\documentclass[11pt]{article}
\usepackage{amsmath,amssymb,bm}
\usepackage{graphicx}

\usepackage{mystyle}
\usepackage{cite,./mcite}
\usepackage{authblk}
\usepackage{hyperref}

\newcommand{\avgcos}{\langle \cos \varphi \rangle}

\newcommand{\veckjone}{{\bf k}_{J,1}}
\newcommand{\veckjtwo}{{\bf k}_{J,2}}

\newcommand{\kina}{$\sqrt{s}=7$ TeV, $|\veckjone|=|\veckjtwo|=35$ GeV}
\newcommand{\kinb}{$\sqrt{s}=14$ TeV, $|\veckjone|=|\veckjtwo|=35$ GeV}
\newcommand{\kinc}{$\sqrt{s}=14$ TeV, $|\veckjone|=|\veckjtwo|=20$ GeV}
\newcommand{\kind}{$\sqrt{s}=14$ TeV, $|\veckjone|=|\veckjtwo|=10$ GeV}

\title{Double scattering contribution to small-x processes: Mueller-Navelet jets at the LHC}
\author[1,2]{B.~Duclou\'e}
\author[3]{L.~Szymanowski}
\author[4,5]{S.~Wallon}
{\tiny
\affil[1]{Department of Physics, University of Jyv\"askyl\"a, P.O. Box 35, 40014 University of Jyv\"askyl\"a, Finland}
\affil[2]{Helsinki Institute of Physics, P.O. Box 64, 00014 University of Helsinki,
Finland}
\affil[3]{National Centre for Nuclear Research (NCBJ), Warsaw, Poland}
\affil[4]{Laboratoire de Physique Th\'eorique, UMR 8627, CNRS,
Univ. Paris Sud, Universit\'e Paris-Saclay, 91405 Orsay, France}
\affil[5]{UPMC Univ. Paris 06, Facult\'e de Physique, 4 place Jussieu, 75252 Paris Cedex 05, France}
}

\begin{document}

\maketitle 

\begin{figure}[h]
	\centering
	\includegraphics[height=4cm]{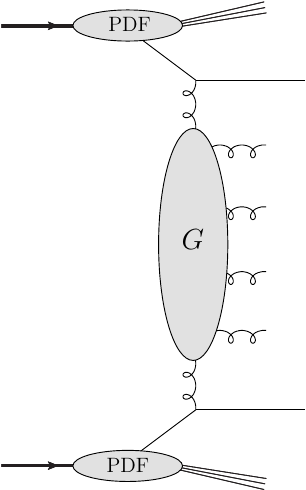}
	\hspace{2cm}
	\centering\includegraphics[height=4cm]{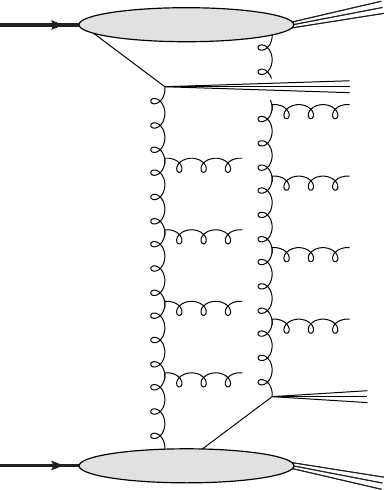}
	\hspace{2cm}
	\includegraphics[height=4cm]{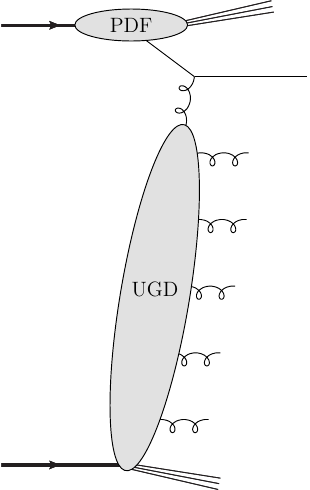}
	\caption{Left: The SPS contribution to Mueller-Navelet jets production at LL accuracy; Middle: The DPS contribution; Right: Inclusive forward jet production.}
	\label{fig:onejet}
\end{figure}

\noindent
The study of the production of two jets separated by a large interval of rapidity at hadron colliders was proposed by Mueller and Navelet~\cite{Mueller:1986ey} to test QCD in the high energy limit, described by the Balitsky-Fadin-Kuraev-Lipatov (BFKL) approach~\cite{Fadin:1975cb,Kuraev:1976ge,Kuraev:1977fs,Balitsky:1978ic}. The standard mechanism for this process is based on single parton scattering (SPS) illustrated in the left panel of Fig.~\ref{fig:onejet} and involves usual parton distribution functions (PDF) of incident partons inside the scattered protons. The classical observable in this process is the azimuthal correlation of the produced jets: in the approach based on collinear factorization of QCD at leading order the two jets would be emitted back-to-back whereas in the BFKL case, with multiple emissions between the two jets described by the BFKL Green's function $G$, one should expect a stronger decorrelation in the azimuthal angle between the two jets. The azimuthal correlations of Mueller-
Navelet jets were measured for the first time at the LHC by the CMS collaboration.  In our recent studies we have shown in~\cite{Ducloue:2013bva} that the measured data on decorrelation coefficients can be well described by a complete next-to-leading logarithmic (NLL) BFKL calculation ~\cite{Colferai:2010wu,Ducloue:2013hia} supplemented by the use of the Brodsky-Lepage-Mackenzie procedure adapted to BFKL dynamics to fix the renormalization scale (for details we refer the reader to~\cite{Ducloue:2013bva}). These results are not severely affected by energy-momentum non-conservation \cite{Ducloue:2014koa}. 

\noindent
However, at high energies and low transverse momenta probed at the LHC, parton densities can become large enough that contributions where two partons from the same incoming hadron take part in the interaction could become important. In the case of Mueller-Navelet jets production such a mechanism is illustrated in the middle panel of Fig.~\ref{fig:onejet}. Thus it is natural to ask about the potential impact of such double parton scattering (DPS) contributions on the good agreement of our earlier results based on single parton scattering (SPS) with experimental data. This was the subject of our recent study  \cite{Ducloue:2015jba}.

\noindent
The main assumption of the analysis in~\cite{Ducloue:2015jba} concerns the independence of the production of each jet. This is the basis of the so-called "pocket formula" for the cross section in the DPS mechanism 
\begin{equation}
	\sigma_{\rm DPS}=\frac{\sigma_{\rm fwd} \sigma_{\rm bwd}}{\sigma_{\rm eff}} \, ,
	\label{eq:sigma_DPS}
\end{equation}
where $\sigma_{\rm fwd (bwd)}$ is the inclusive cross section for one jet in the forward (backward) direction as shown in the right panel of Fig.~\ref{fig:onejet} and $\sigma_{\rm eff}$ is a phenomenological quantity related to the density of the proton in the transverse plane. According to measurements at the Tevatron and the LHC, $\sigma_{\rm eff}$ should be of the order of 15 mb but there is some discrepancy between these measurements. To account for this uncertainty we vary $\sigma_{\rm eff}$ between 10 and 20 mb in our calculation.

\noindent
A more refined approach to compute the DPS contribution shown in the middle panel of Fig.~\ref{fig:onejet} would require to introduce some kind of ``hybrid'' double parton distribution, related to the probability for a proton to emit both a collinear parton and a gluon with some transverse momentum. Since almost nothing is known about such distributions we are forced to use the simplistic factorized model~(\ref{eq:sigma_DPS}).

\noindent
The cross section for one forward jet production, $\sigma_{\rm fwd}$, shown in the right panel of Fig.~\ref{fig:onejet} requires the knowledge of the unintegrated gluon density (UGD). Contrary to usual collinear parton distributions (PDF), the unintegrated gluon distributions are not very well known. To estimate the uncertainty of our calculation due to this fact, we will use several sets found in the literature to see the impact of this choice on our final results. For each set we fix the normalization by a comparison with CMS data on inclusive forward jet production (for more details see~\cite{Ducloue:2015jba}).
\begin{figure}[h!]
	\centering
	\includegraphics[width=11cm]{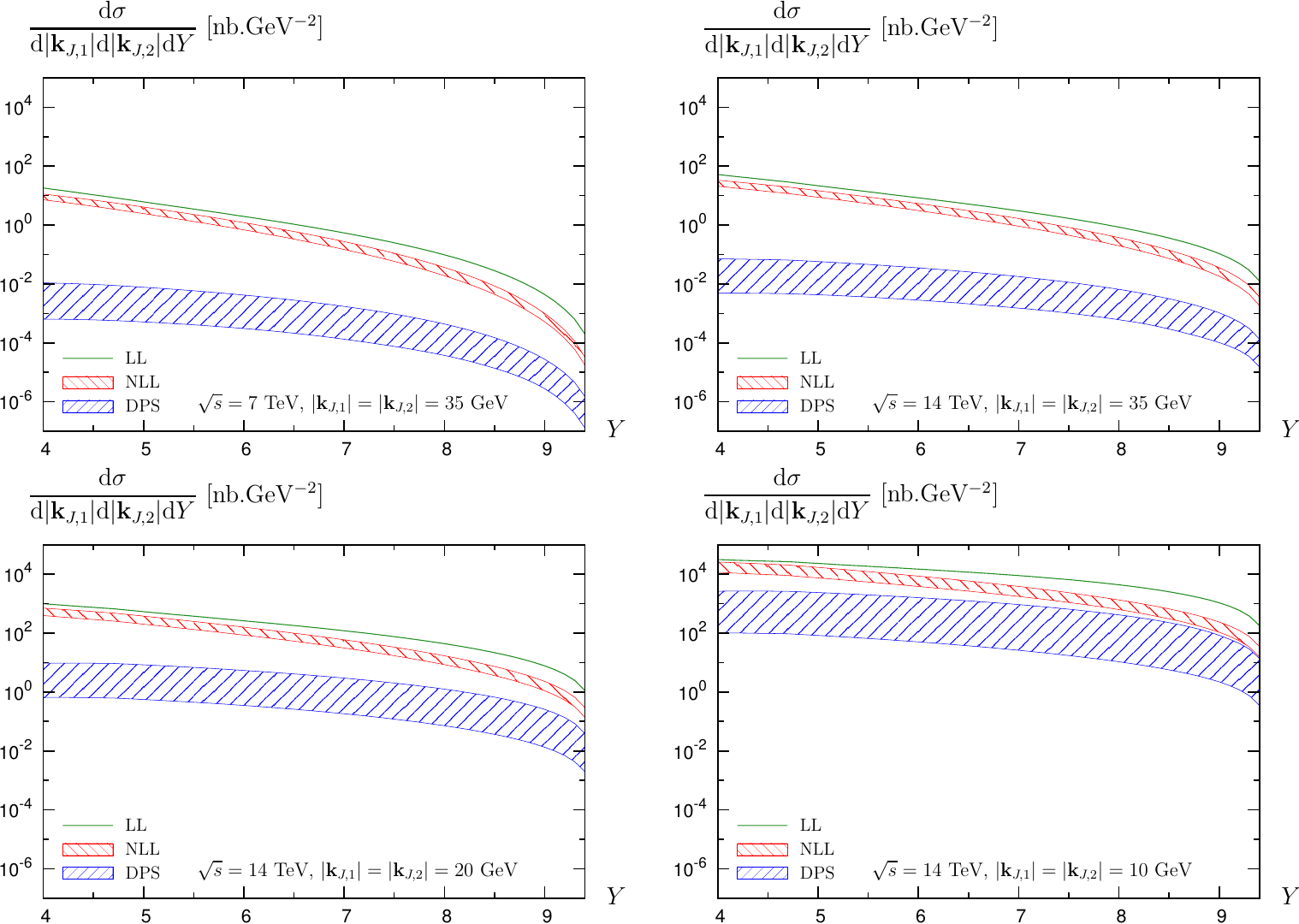}
	\caption{Comparison of the differential cross section obtained at LL (green) and NLL (red) accuracy in the BFKL approach and the DPS cross section (blue) for the four kinematical cuts described in the text.}
	\label{fig:sigma}
\end{figure}

\begin{figure}[t]
	\centering
	\vspace{-.9cm}\includegraphics*[width=13.2cm,angle=0]{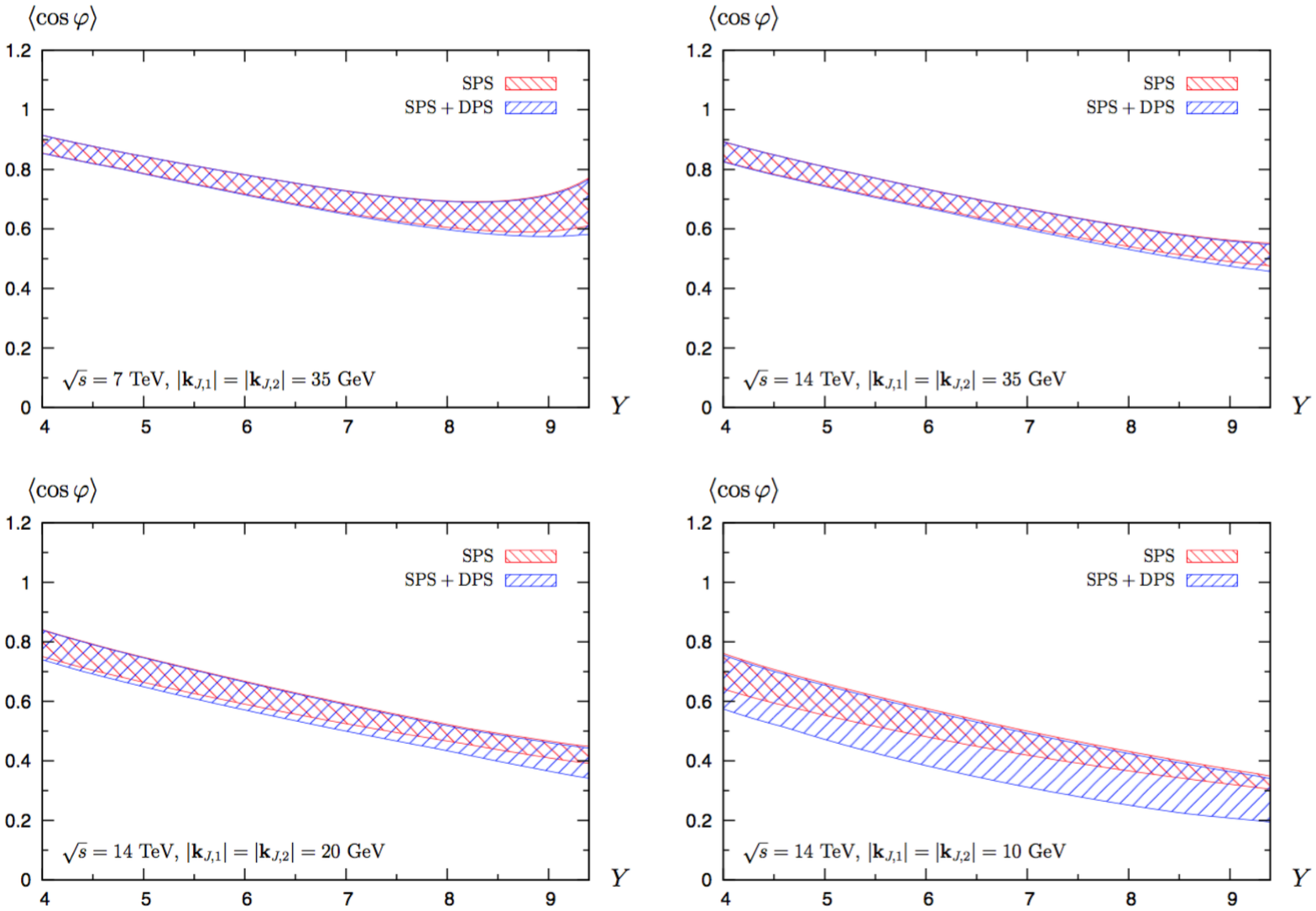}	
	\vspace{-.9cm}
	\caption{$\avgcos$ as a function of $Y$ without (red) and with (blue) the DPS contribution for the four kinematical cuts described in the text.}
	\label{fig:cos}
\end{figure}

\noindent
Our analysis about the importance of DPS in Mueller-Navelet jets production is performed for four choices of kinematical cuts on $\sqrt{s}$ and the transverse momenta of the jets relevant for present and forthcoming experiments at the LHC:
\begin{itemize}
	\item \kina,
	\item \kinb,
	\item \kinc,
	\item \kind.
\end{itemize}
The resulting predictions for the differential cross section and the decorrelation coefficient $\avgcos$ are shown on Figs.~\ref{fig:sigma} and~\ref{fig:cos} respectively. One can see that in most cases the DPS cross section is much smaller than the SPS one and thus the impact of this contribution on the azimuthal correlation is rather small.
However, if one considers the set of parameters giving the largest DPS contribution, for low transverse momenta and large rapidity separations the effect of DPS can become larger than the uncertainty on the NLL BFKL calculation. Therefore in this region a more careful analysis or experimental data would be required to conclude. 
  

\section*{Acknowledgements}
This work is supported by
French grant ANR PARTONS No. 
ANR-12-MONU-0008-01.
B. D. is supported by the Academy of Finland, project 273464.
L. S. was partially supported by grant of National Science Center, Poland, No. 2015/17/B/ST2/01838. 
This work was done using computing resources from CSC -- IT Center for Science in Espoo, Finland.

\end{document}